\documentstyle[aps,psfig,eqsecnum]{revtex}
\begin{document}
\twocolumn

\def\beq{\begin{equation}}
\def\eeq{\end{equation}}
\def\la{\langle}
\def\ra{\rangle}
\def\ep{e^{\pi \omega/2 \kappa}}
\def\em{e^{-\pi \omega/2 \kappa}}
\def\epm{e^{\pm \pi \omega/2 \kappa}}
\def\emp{e^{\mp \pi \omega/2 \kappa}}
\def\ept{e^{\pi \omega/ \kappa}}
\def\emt{e^{-\pi \omega/\kappa}}
\def\o{\omega}
\def\k{\kappa}
\def\Arg{{\rm Arg}}

\draft

\title{Black hole lasers} 

\author{Steven Corley\thanks{scorley@phys.ualberta.ca}}

\address{Theoretical Physics Institute,
Department of Physics, University of Alberta,
Edmonton, Alberta, Canada T6G 2J1}

\author{Ted Jacobson\thanks{jacobson@physics.umd.edu}} 
\address{Department of Physics, University of Maryland,
College Park, MD 20742-4111, USA}

\maketitle
\begin{abstract}
High frequency dispersion does not alter the low frequency 
spectrum of Hawking radiation from a single black hole horizon, 
whether the dispersion entails subluminal or superluminal group 
velocities. We show here that in the presence of an inner horizon 
as well as an outer horizon the superluminal case differs dramatically 
however. The negative energy partners of Hawking quanta return to the 
outer horizon and stimulate more Hawking radiation if the field is 
bosonic or suppress it if the field is fermionic. This process leads 
to exponential growth or damping of the radiated flux and correlations 
among the quanta emitted at different times, unlike in the usual 
Hawking effect. These phenomena may be observable in condensed matter 
black hole analogs that exhibit ``superluminal" dispersion.
\end{abstract}

\section{Introduction} 

Recent work has shown that Hawking radiation is highly 
insensitive to modifications of the short
distance physics of the quantum field.
In these models linear fields are considered, and the
field equation is modified at high wavevectors in 
some preferred frame, yielding a nonlinear dispersion
relation $\omega(k)$ relating frequency to wavevector.
Models with both subluminal\cite{Unruh95,BMPS,CJ} and 
superluminal\cite{Unruhsuper,Corsuper} group velocities
at high wavevectors have been studied, including lattice black
hole spacetimes\cite{CJlattice} (which have subluminal dispersion).
The picture that emerges from these studies is that 
the thermal Hawking spectrum is very robust for black
holes with temperature much less than the energy scale 
of the new physics. Although the short distance physics
does modify this spectrum, the modifications are so slight
at the frequencies of interest that they seem well nigh 
impossible to observe.  

We have found a dramatic exception to this rule however.
If there is both an outer and an inner horizon, 
and if the dispersion is superluminal, then the Hawking 
process for a bosonic field 
is self-amplifying and the radiated flux grows
exponentially in time, while for a fermionic field the process is 
self-attenuating. What happens is that the negative
energy partner of a Hawking particle, after falling to 
the inner horizon, ``bounces" and returns to the outer
horizon on a superluminal trajectory, where it 
either stimulates or suppresses
more Hawking radiation in the bosonic or fermionic case respectively. 
This secondary radiation is not
only different than the usual Hawking flux, but it is
correlated to the prior radiation. In the bosonic case 
the process continues
to amplify at least until the back reaction becomes important. 

Charged black holes have inner horizons, but astrophysical
ones would loose their charge very rapidly, so it is difficult
to imagine how this runaway Hawking effect could ever be 
observed for real black holes. Even so, it provides an 
interesting theoretical laboratory in which to explore the 
effects of short distance physics. Moreover, it is conceivably
relevant to string theory, and it might be observable in a
condensed matter analogue of a black hole. Let us briefly
indicate these ideas in turn.

In spite of many points of close agreement between the physics
of near extremal D-branes and black holes, a glaring 
discrepancy persists. If a radiating near extremal D-brane state 
is maintained at fixed energy by a constant influx of energy
in a pure state, then the entropy in the radiation will be 
constant and there will be correlations in the 
radiation that emerges at different times. For a black hole,
on the other hand, the usual Hawking process leads to uncorrelated
thermal radiation for all time. The effects of superluminal 
dispersion invalidate the usual Hawking picture because the 
negative energy partners return to the event horizon. 
If there is something analogous to the superluminal dispersion of our
model in string theory, then perhaps that could eliminate the 
discrepancy between the string and black hole pictures. 
This may not be so far-fetched. String theory is, after all, 
non-local in some sense, and there is some evidence\cite{Lowe} 
suggesting that it supports superluminal effects. 

A condensed matter analog---Unruh's sonic black
hole\cite{Unruh81,Unruh95,Visser}---was
the original stimulus for the development of the dispersive
models. In this model, a sonic horizon occurs where the flow 
velocity of an inhomogeneous fluid exceeds the speed of sound.
Although it seems unlikely that this situation can be experimentally
realized for a low temperature superfluid, there are variations
of the idea that might be realizable, involving quasiparticles
other than phonons in different systems. For example, 
this may occur for fermion quasiparticles in rotating superfluid 
vortex cores with gap nodes 
such as $^3$He-A or $d$-wave superconductors\cite{KopnVolo}, or in moving
$^3$He-A textures\cite{JacoVolo}. In both these examples there are both
inner and outer horizons. Moreover, the quasiparticle
dispersion relation is ``relativistic" sufficiently near a
gap node, and the group velocity increases (i.e. becomes ``superluminal")
as the difference between the momentum and the gap node increases,
so the effective field theory
has ``superluminal" dispersion. Thus it is not inconceivable
that the phenomena discussed here may someday be observable.

This paper is organized as follows.
In section \ref{model} the superluminal dispersion model 
for both bosons and fermions is discussed. 
The propagation of wavepackets in the black
hole spacetime with inner and outer horizons is analyzed
qualitatively in section \ref{wp} and the implications for 
the amplification or suppression and the correlations
in the Hawking radiation
are drawn in section \ref{creation}. Section \ref{quantitative}
renders the previous discussion quantitative by using explicit
wavepacket solutions (derived in the appendix) to find
expressions for the number and
correlations between the radiated quanta. 
Open issues concerning
the boundary conditions on the quantum state and the 
gravitational back reaction
are discussed briefly in section \ref{discussion}.
 
We use units with $\hbar=c=1$ and metric signature ($+$$-$$-$$-$).

\section{Superluminal dispersion model}
\label{model}
A 2-dimensional model suffices to illustrate the essential physics.
We assume the spacetime metric is static, and therefore\cite{Painleve}
coordinates can be chosen (at least locally)
so that the line element takes the form
\begin{equation}
ds^2 = dt^2 - \bigl(dx - v(x) dt\bigr)^2.
\label{metric}
\end{equation}
A special case is the line element of the $t-r$ subspace of the 
Reissner-Nordstr\"om black hole spacetime in Painlev\'e-Gullstrand 
coordinates, where $v(r) = - \sqrt{2 GM/r - Q^2/r^2}$. (These coordinates
cover the black hole interior down to where $v(r)=0$, at $r=Q^2/2GM$.)
More generally, we consider any $v(x)$ which is negative,
vanishes as $x\rightarrow +\infty$, and is greater than $-1$
except between inner and outer horizons, located 
at $x_i$ and $x_o$, where $v(x_{i,o})=-1$.

\subsection{Boson field} 
\label{boson}
We adopt a linear field theory with higher spatial derivatives
included in the action in order to provide a superluminal
dispersion relation. In this section we restrict to the
case of a real bosonic field. The case of a Majorana fermion 
field will be discussed in section \ref{fermion}. 
The action for the field is given by
\begin{equation}
S_\phi = \frac{1}{2} \int d^2 x \, \left[ \Bigl( (\partial_t 
+ v \partial_x) \phi \Bigr)^2
+ \phi {\widehat F}(\partial_x) \phi \right].
\label{action}
\end{equation}
In the ordinary relativistic action one has ${\widehat F}(\partial_x) = 
\partial_{x}^{2}$.  In this paper we take
\begin{equation}
{\widehat F}(\partial_x) = \partial^{2}_x - \frac{1}{k_{0}^{2}}
\partial^{4}_x.
\label{derivop}
\end{equation}
To motivate this action we note that the black hole defines a preferred
frame, the frame of freely falling observers.  In the Painlev\'e-Gullstrand
coordinate system, $(\partial_t + v \partial_x)$ is the unit tangent to 
free fall world lines that start from rest at infinity, and $\partial_x$ is
its unit, outward pointing normal. Our action comes from modifying
the derivative operator, only along the unit normal $\partial_x$, by the
addition of higher derivative terms which become important
only when the wavelength is of order $1/k_0$ or shorter. We will
assume that this length scale of ``new physics" is much
shorter than the length scale of the metric (\ref{metric}), 
i.e. $k_0\gg |v'/v|$. (In particular, we assume $k_0\gg\kappa$, where
$\kappa=|v'(x_{i,o})|$ is the surface gravity of the horizon.) 
The idea is that the microstructure
of spacetime, or of a condensed matter analog, might give rise to 
such higher derivative terms in the effective action. 
The choice (\ref{derivop}) is just
the generic form for the lowest order such term that is reflection
invariant and produces superluminal group velocities.

The action (\ref{action},\ref{derivop})  
produces the equation of motion
\begin{equation}
(\partial_t+\partial_x v)(\partial_t+v\partial_x)\phi=
\partial_{x}^{2} \phi - \frac{1}{k_{0}^{2}} \partial_{x}^{4} \phi.
\label{eom}
\end{equation}
To derive the dispersion relation for this equation we 
look for solutions of the form
\begin{equation}
\phi(t,x) = \exp \left(-i \omega t +i\int^{x} k(x') \, dx' \right)
\end{equation}
where $k(x)$ is a position dependent wavevector.  Substituting
this ansatz into the equation of motion (\ref{eom}) and neglecting
derivatives of $v(x)$ and $k(x)$ results in the dispersion relation
\begin{equation}
 (\omega - v k )^2  = F^2 (k)
\label{dispreln}
\end{equation}
where
\begin{equation}
F^2(k) = k^2 + k^4/k_{0}^{2}.
\label{F}
\end{equation}
The group velocity in the free-fall frame is $dF/dk$,
so wavepackets with $k \ll k_0$ 
propagate near the speed of light, whereas  
wavepackets with $k\gtrsim k_0$ propagate superluminally.

\begin{figure}[hbt]
\centerline{
\psfig{figure=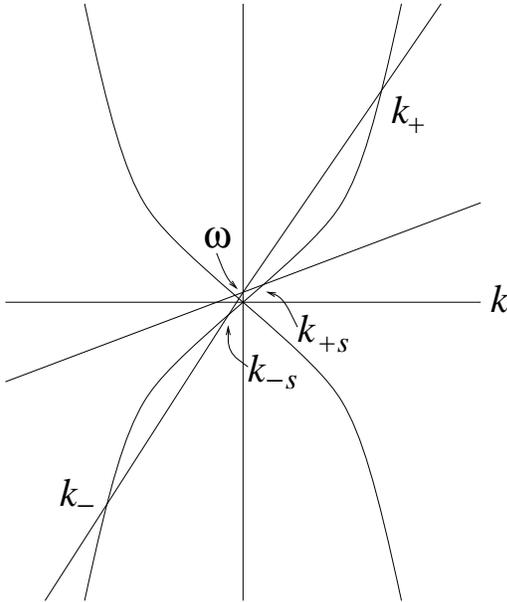,angle=-90,height=8cm}}
\caption{\small Plot of $(\omega + |v| k)$ (for one value of 
$\omega$ and two values
of $v$) and $F(k)$ as functions of $k$.  
The intersection points of the curves are the
allowed wavevector roots of the dispersion relation (\ref{dispreln}).} 
\label{dispgraph}
\end{figure}

The dispersion relation (\ref{dispreln}) is a fourth order polynomial 
equation in the wavevector $k$ so it has four solutions
for $k$ at given values of $\omega$ and $v$.
The nature of these roots is revealed by a graphical method.
In figure \ref{dispgraph} we plot the straight line
$(\omega + |v| k)$ for one value of $\omega$ 
(satisfying $0<\omega\ll k_0$) and two values of $v$, 
and the curve $\pm F(k)$, as functions of $k$. (We define $F(k)$
as the {\it positive} square root of (\ref{F}).)
The intersection points are the allowed real
wavevector roots to the dispersion relation.
When $|v| <1$ there are only two $real$  roots (corresponding
to the two roots for the ordinary dispersion relation with
$F_{ord}(k) = \pm k$), the other 
two being complex.  The positive wavevector is denoted $k_{+s}$.  
When 
$|v| \gtrsim 1+ \frac{3}{2}(\omega/k_0)^{2/3}\approx 1$,
on the other hand, all four roots are real, with one positive and
three negative.
The positive wavevector is denoted
$k_+$ in this case and, in decreasing magnitude, the negative wavevectors
are denoted $k_-$ and $k_{-s}$ respectively (the other negative
wavevector corresponds to an ingoing wave that  
plays no role in this paper so we do not give it a name). 
These roots are labeled in figure \ref{dispgraph}.
 
The dispersion relation plot in figure \ref{dispgraph} is also quite 
convenient for tracing the motion of wavepackets
in the background spacetime. The coordinate group velocity 
$v_g=dx/dt$ 
of a wavepacket centered on a given wavevector is given by 
\beq
v_g=\frac{d\omega}{dk}=-|v|\pm\frac{dF}{dk},
\label{vg}
\eeq
where $\pm dF/dk$ is the group velocity in the free-fall frame.
Thus at any wavevector $v_g$ is just the slope of the $\pm F(k)$ 
curve minus the slope of the straight line $(\omega + |v| k)$.
For all four types of wavevectors $k_{\pm s, \pm}$ of interest to us, 
$\pm dF/dk$ is positive, hence the sign of $v_g$ is determined by 
which of the two slopes is larger, something that 
is easily read from the figure. For $\omega>0$, the group
velocity for $k_{+s}$ and $k_\pm$ is positive, whereas for 
$k_{-s}$ it is negative.

When generalized to a complex scalar field, 
the action (\ref{action}) is invariant under constant
phase transformations of the field.
This implies the existence of a conserved current
$j^{\mu}$.  The integral of the time component $j^{0}$ over a spatial
slice serves as a conserved inner product when evaluated on complex
solutions to the equation of motion (\ref{eom}).  
For the metric (\ref{metric}), this inner product takes the form
\begin{equation}
(f,g) = i \int dx\,  \Bigl(f^{*}
(\partial_{t} + v\partial_{x})g - g(\partial_{t} + v\partial_{x})
f^{*}\Bigr),
\label{inner}
\end{equation}
where $f(t,x)$ and $g(t,x)$ are solutions to (\ref{eom}).

Two classes of complex 
solutions to the field equation (\ref{eom}) are of special interest
for quantization.
The first are the positive free fall
frequency wavepackets.  They
can be written as sums of solutions satisfying
\begin{equation}
(\partial_{t} + v \partial_{x})f(t,x) = -i \omega^{\prime}f(t,x)
\end{equation}
where $\omega^{\prime} >0$.
The second are the positive Killing frequency wavepackets.
These are sums of solutions of the form
$e^{-i \omega t} \varphi(x)$ where $\omega > 0$.
A positive free fall frequency wavepacket 
confined to a constant $v(x)$  interval at one time 
necessarily has a positive
norm under (\ref{inner}), as does a positive Killing frequency wavepacket 
confined to a region where $v(x) = 0$ (where Killing
frequency coincides with free-fall frequency). Since the norm is
conserved, it is  positive at all times if it is at one 
time, even when the wavepacket does not remain in an interval
of constant or vanishing $v(x)$. Note that 
if the wavelength is small compared
to the scale of variations of $v(x)$, then a positive free-fall
frequency wavepacket will have positive norm even if $v(x)$ is
not constant. 
 
To quantize the field we assume that $\widehat\phi(t,x)$ is a self-adjoint
operator solution to the field equation that satisfies the
canonical commutation relations.  We define the annihilation operator 
$a(f)$ associated to a normalized complex solution to the wave equation
$f(t,x)$ by
\begin{equation}
a(f) \equiv (f,\widehat\phi).
\label{annih}
\end{equation}
The commutation relations for the field operator are equivalent 
to the  relations 
\beq
[a(f),a^\dagger(g)]=(f,g)
\label{ccr}
\eeq
for all $f$ and $g$.
If $f(t,x)$ is a positive norm solution, then $a(f)$ behaves 
as an annihilation operator. If $f(t,x)$ is a negative
norm solution, $f^*(t,x)$ has positive norm, so 
$a(f)= -a^\dagger(f^*)$ behaves as a creation operator.

\subsection{Fermion field}
\label{fermion}

For simplicity we consider two-dimensional massless Majorana fermions.
Following the conventions of \cite{GSW}, the action 
in a general curved spacetime is given by\footnote{In
higher dimensions there would be a spin connection term as well.
In two-dimensions it is easy to show that this term vanishes
identically.} 
\begin{equation}
S_{\psi} = \frac{i}{2} \int d^2 x \sqrt{-g} \bar{\psi} \Gamma^{\mu}
\partial_{\mu} \psi
\label{fermaction}
\end{equation}
where $\Gamma^{\mu} = \Gamma^{a} e_{a}^{\mu}$ and $e_{a}^{\mu}$
is the zweibein.  We take the flat space gamma matrices as
\begin{equation}
\Gamma^{0} = \left( \begin{array}{cc}
0  &  -i \\
i  &  0
\end{array} \right), \,\,
\Gamma^{1} = \left(\begin{array}{cc}
0  &  i  \\
i  &  0
\end{array} \right).
\end{equation}
Decomposing the spinor $\psi$ as
\begin{equation}
\psi = \left( \begin{array}{c}
\psi_{+} \\ \psi_{-} \end{array} \right)
\end{equation}
and expanding the action in the metric (\ref{metric})
using the zweibein 
$(e_0,e_1)=(\partial_t+v\partial_x,\, \partial_x)$ we find
\begin{eqnarray}
S_{\psi} = \frac{i}{2} \int d^2 x 
\Bigl ( & \psi_+ & \bigl(\partial_t + (1+ v)\partial_x \bigr) \psi_+ 
\nonumber \\
+ & \psi_- & \bigl(\partial_t -(1- v) \partial_x \bigr)\psi_-  \Bigr).
\end{eqnarray}
In this form it is clear that $\psi_+$ and $\psi_-$ do not mix.  
Furthermore at infinity, where $v(x) = 0$, $\psi_+$ is right-moving
while $\psi_-$ is left-moving.  We therefore drop $\psi_-$ in the
remainder as it plays no role in the Hawking radiation calculation.

Following the same motivation described in section \ref{model},
we now modify the action for $\psi_+$ by subtracting the
higher derivative term $k_0^{-2}\psi_+ \partial^{3}_x \psi_+$, obtaining
the action  
\begin{equation}
S_{\psi} = \frac{i}{2} \int d^2 x \left( \psi_+ (\partial_t + (1+ v)
\partial_x - k_0^{-2}\partial^{3}_x) \psi_+ \right). 
\label{faction}
\end{equation}
Varying with respect to $\psi_+$ results in the equation of motion
\begin{equation}
\Bigl(\partial_t + v \partial_x + \partial_x v/2 + \partial_x
- k_0^{-2}\partial^{3}_x\Bigr) \psi_{+} = 0.
\label{feom}
\end{equation}
Substituting
$\psi_+(t,x) = \exp(-i \omega t +
i \int^x k(x')dx')$ into the equation of motion and dropping
derivatives of $k(x)$ and $v(x)$ results in the dispersion
relation 
\begin{equation}
\omega - v  k = k + k^3/k_0^2.
\end{equation}
This is the same 
(up to the coefficient of the $k^3$ term and higher order terms) 
as the branch of the scalar field dispersion relation corresponding to
positive group velocity in the free-fall frame
given in (\ref{dispreln}) and displayed in figure \ref{dispgraph}.  
The classification of scalar wavepacket types in  
section (\ref{boson})
therefore applies to fermion wavepackets as well.
In particular, the higher derivative term leads to 
superluminal propagation at large wavevectors. 
 
To quantize the field we assume that $\widehat\psi_+(t,x)$ is a self-adjoint
operator solution to the field equation that satisfies the
canonical anti-commutation relations 
$\{\widehat\psi_+(t,x),\widehat\psi_+(t,x')\}=\delta(x,x')\}$.
The conserved inner product is the integral of 
the time component of the conserved current
associated with phase invariance of the action (\ref{faction}) (generalized
to complex fermions), and takes the form 
\begin{equation}
\langle \psi_1, \psi_2 \rangle = \int dx\, \psi_{1}^* \psi_2.
\label{fip}
\end{equation}
We define the annihilation operator 
$b(f)$ associated to a normalized complex solution to the wave equation
$f(t,x)$ by
\begin{equation}
b(f) \equiv \la f,\widehat\psi_+\ra.
\label{fannih}
\end{equation}
The anti-commutation relations for the field operator are then equivalent 
to the relations 
\beq
\{b(f),b^\dagger(g)\}=\la f,g\ra
\label{fccr}
\eeq
for all $f$ and $g$.
We represent the operators $b(f)$ on the fermionic Fock space generated 
by positive free-fall frequency 
solutions to the equation of motion (\ref{feom}).  
If $f(t,x)$ is a positive free-fall frequency solution then $b(f)$ behaves 
as an annihilation operator on this space. If $f(t,x)$ is a negative
free-fall frequency solution, then $f^*(t,x)$ has positive free-fall 
frequency, so $b(f)= b^\dagger(f^*)$ behaves as a creation operator.

\section{Wavepacket propagation}
\label{wp}

In this section we give a qualitative analysis of the 
role of the inner horizon in modifying the Hawking radiation.
This analysis will exploit a WKB description of wavepacket
propagation, allowing for non-WKB ``mode conversion" in the 
vicinity of the horizons. The analysis applies equally well
for the bosonic and fermionic quantum fields. 
Scattering of waves on account of the background curvature
of the metric (\ref{metric}) is negligible as long as
the radius of curvature is much greater than $1/k_0$.  
For small wavevectors, $k\ll k_0$, this is because
the wave equation is approximately
conformally invariant and the metric (like any two-dimensional
metric) is conformally flat. For large wavevectors, 
$k\gtrsim k_0$, it is because the wavelength
is much smaller than the radius of curvature.  

\begin{figure}[hbt]
\centerline{
\psfig{figure=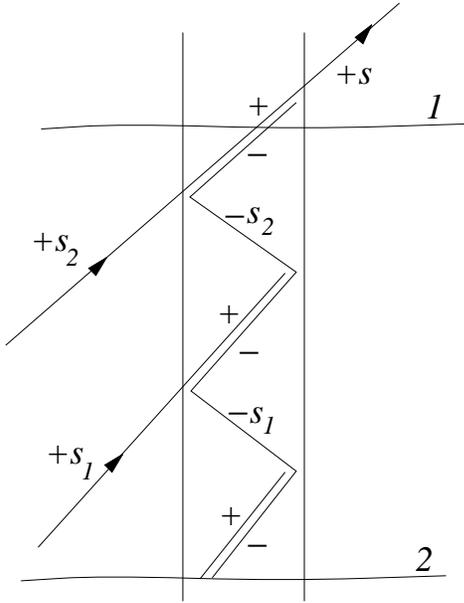,angle=-90,height=8cm}}
\caption{\small Spacetime sketch of the evolution of an outgoing
$k_{+s}$ wavepacket backward in time. 
A line end indicates a wavepacket arising from 
mode conversion, while an unbroken line indicates that the wavevector
evolves continuously on the dispersion curve. 
} 
\label{instabgraph}
\end{figure}

We begin outside the outer horizon with a low frequency outgoing 
wavepacket peaked around a wavevector of type
$k_{+s}$ (see figure \ref{dispgraph}), and we follow this wavepacket
backwards in time. A sketch of what we find is given in 
figure \ref{instabgraph}. The final wavepacket (i.e. the one we begin 
with) is labeled $+s$ in figure \ref{instabgraph}.  
This packet has positive group
velocity and therefore is right-moving, as can be
seen from the graph of the dispersion relation 
(Fig. \ref{dispgraph}).
Following this packet {\rm backward} in time it moves toward the black
hole and blueshifts. The Killing frequency $\omega$ is conserved,
so the increase in the wavevector can be seen from Fig. \ref{dispgraph}
by increasing the slope of the straight line while keeping the intercept 
fixed.  As the wavevector grows, the group velocity  
increases in the free-fall frame, 
and so the packet becomes superluminal and crosses the horizon 
(backward in time), 
becoming a packet with wavevectors of type $k_+$ 
(see figure \ref{instabgraph}).  

The wavepacket inside the horizon also has 
a $k_-$ component, which is not obvious if we 
simply follow continuously along the dispersion
curve. In fact, the WKB approximation
breaks down near the horizon, and ``mode conversion"
from the positive wavevector to the negative wavevector, 
negative free-fall frequency, branch 
of the dispersion relation occurs. 
This is easily shown analytically, and is made plausible
by the fact that,  around the horizon, 
the straight line of figure \ref{dispgraph} 
nearly coincides with a large portion of the curved line of the 
dispersion curve, thus allowing other wavevectors to become
mixed in. The dispersion relation allows wavevectors 
of types $k_+$, $k_-$, and $k_{-s}$ in between the horizons,
however only the first two are right moving, whereas the
last type is left moving. Since our final wavepacket is by
assumption purely outgoing outside the horizon, there can be no
$k_{-s}$ component generated here. The $k_+$ and $k_-$ wavepackets
are labeled $+$ and $-$ in figure \ref{instabgraph}.
In this figure a line end indicates a wavepacket arising from 
mode conversion, while an unbroken line indicates that the wavevector
evolves continuously on the dispersion curve.
 
The $k_+$ and $k_-$ packets propagate backward in time toward
the inner horizon where they both undergo partial mode conversion.  
The group velocity of the $k_+$ packet
remains positive around the inner horizon and therefore it can cross,
becoming a $k_{+s}$ packet, labeled 
$+s_2$ in figure \ref{instabgraph}.  
As before though, there is also some mode
conversion from the positive to the negative wavevector branch
of the dispersion relation, and a left-moving $k_{-s}$ packet 
($-s_2$ in figure \ref{instabgraph}) is generated 
which propagates backward in time back toward the
outer horizon.  The $k_-$ packet on the other hand cannot cross
the inner horizon on the negative wavevector branch 
because its group velocity drops to zero at the horizon. 
Indeed the group velocity goes through zero and becomes negative,
so the $k_-$ packet turns around and propagates
back toward the outer horizon as a $k_{-s}$ packet still on the negative
wavevector branch.  In addition, some mode conversion from the negative
to the positive wavevector branch of the dispersion relation occurs
at the inner horizon.
Therefore part of the $k_-$ packet does cross the horizon
as a $k_{+s}$ packet and is superposed with the $k_{+s}$
packet that evolved from the $k_+$ packet. 
 
The $k_{+s}$ packet inside the inner horizon continues
propagating to the left backward in time.  The $k_{-s}$
packet however returns to the outer horizon, near which
its group velocity drops to zero. Again, partial mode conversion
to the positive wavevector branch occurs, so the 
$k_{-s}$ packet evolves backward in time
to a pair of $k_+$ and $k_-$ packets which are heading back to
the inner horizon.  This is now almost the same situation
we started with, since the original $k_{+s}$ packet also
evolved into a pair of $k_{+}$ and
$k_-$ packets between the horizons (although with 
a different relative weight). The analysis 
given above thus tells us qualitatively what happens
when they reach the inner horizon, namely, the same thing
as happened before. The general pattern that emerges is shown
in figure \ref{instabgraph}.

\begin{figure}[hbt]
\centerline{
\psfig{figure=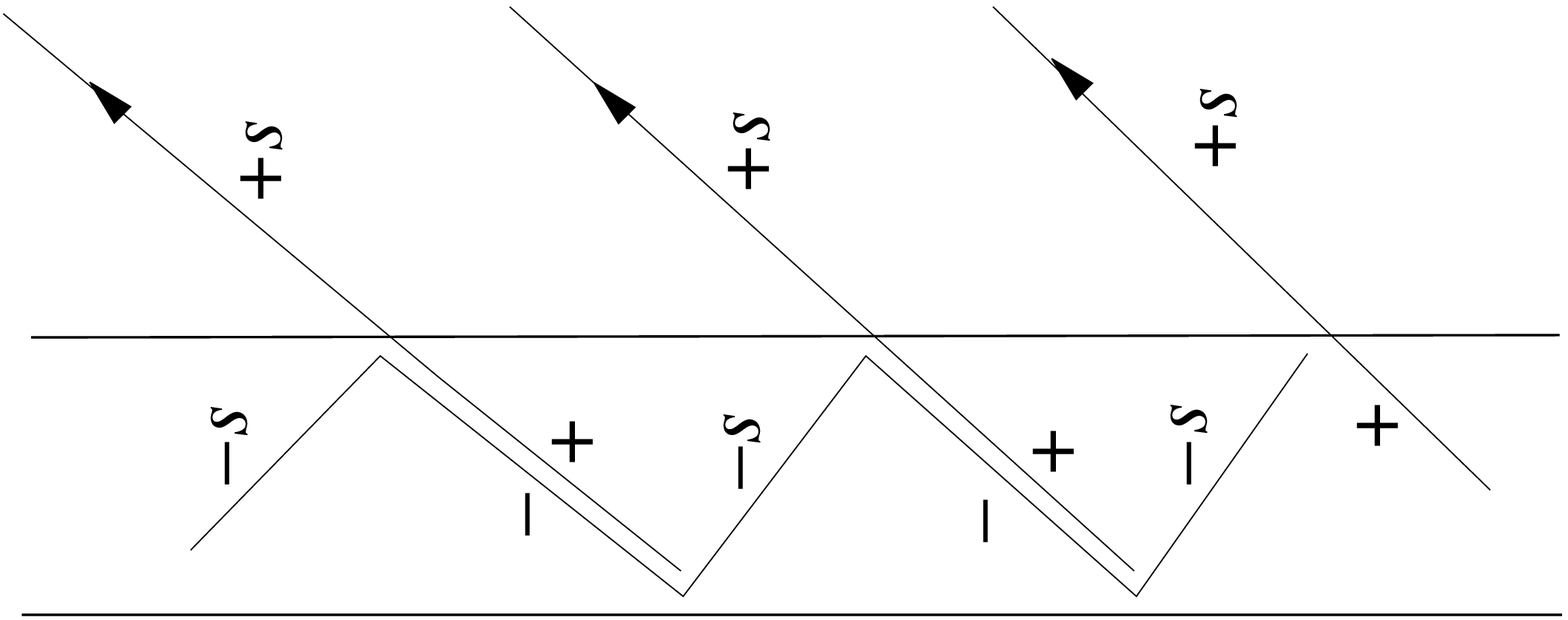,angle=-90,height=8cm}}
\caption{\small Spacetime sketch of the trajectory of a
Hawking particle and its partner forward in time.  
A line end indicates a wavepacket arising from 
mode conversion, while an unbroken line indicates that the wavevector
evolves continuously on the dispersion curve.
} 
\label{partner}
\end{figure}

We have so far discussed the history of an outgoing 
$k_{+s}$ wavepacket followed backward in time.
It is also instructive to look at the {\it future} evolution
of a $k_{-s}$ wavepacket in between the horizons, since
the negative energy partner of a Hawking particle is such
a wavepacket. This evolution can be inferred by the same
sort of analysis just given, or simply by time and space reversal
of that analysis, and is shown in figure \ref{partner}.

\section{Particle creation: origin of the amplification
or suppression of Hawking radiation}
\label{creation}
The amount of particle creation in an outgoing 
positive frequency wavepacket $\psi$ is indicated by 
the expectation value of the number operator 
$N(\psi)=a^\dagger(\psi)a(\psi)$. To determine this expectation value
an initial quantum state must be specified.  Let us define
an in-Hilbert space on some spacelike surface as the Fock
space generated by positive free-fall frequency wavepackets
on that surface. The corresponding ground state is then annihilated
by annihilation operators of these wavepackets. We shall suppose
the initial state is such a free-fall ground state associated with 
a given surface $\Sigma$. Decomposing $\psi=\psi^++\psi^-$ into its
positive and negative free-fall frequency parts on $\Sigma$,
the ground state condition implies in the bosonic case that 
$\la N(\psi)\ra=-(\psi^-,\psi^-)$, and in the fermionic case
$\la N(\psi)\ra=\la\psi^-,\psi^-\ra$. 

Suppose we choose $\Sigma$ as surface 1 in
figure \ref{instabgraph}, i.e., a surface that cuts through the $k_+$ and
$k_-$ packets first produced by propagating the $k_{+s}$ packet back
in time. Then the number expectation value for the $k_{+s}$
packet is just (minus) the norm of the $k_-$ packet. 
In \cite{Unruhsuper,Corsuper} this was shown (for bosons) to be thermal 
at the Hawking temperature, for wavepackets with Killing
frequencies $\omega$ satisfying $\kappa\lesssim\omega\ll k_0$.
That is, the standard Hawking effect occurs
even in the presence of superluminal dispersion, if there is
only a single horizon. 

When there is also an inner horizon,
the particle creation depends very much on which surface is used
to define the initial ground state.
If we impose the ground state condition on the earlier surface 
2 in figure \ref{instabgraph}, instead of surface 1,
the occupation number for the final $k_{+s}$ packet is no longer
thermal. The norm of the negative frequency part 
of the wavepacket on surface 2 is determined not just by 
the final passage across the outer horizon, but also by 
the mode conversion processes at the inner and outer horizons.

As the time between the initial ground state and the
final outgoing wavepacket grows, there is an exponential
amplification or suppression in the occupation number of 
the final wavepacket in the boson and fermion cases respectively.
To see why, note that 
the $k_{-s}$ packet denoted $-s_2$ in figure \ref{instabgraph} evolves into
the orthogonal $k_{+s}$ and $k_{-s}$ packets 
denoted $+s_1$ and $-s_1$ respectively, hence the norms are related 
by 
\beq
||-\!s_2||^2  = ||+\!s_1||^2 +||-\!s_1||^2 
\label{amp}
\eeq
where $||f||^2$ stands for 
$(f,f)$ in the bosonic case and $\la f,f\ra$ in the fermionic case.

Consider first the bosonic case.
A $k_{+s}(k_{-s})$ packet has positive (negative) 
free fall frequency 
and therefore positive (negative) norm under (\ref{inner}),
so it follows from (\ref{amp}) 
that $||-\!s_1||^2$ is larger in magnitude than $||-\!s_2||^2$.
Continuing into the past this process repeats,
and for each ``bounce" between the horizons the norm
of the wavepacket between the horizons grows
by some fixed multiple, resulting in exponential
growth of both $||-\!s_n||^2$ and $||+\!s_n||^2$.\footnote{It
is perhaps surprising to have exponential growth in time
when there are no imaginary frequency solutions to the 
dispersion relation (\ref{dispreln}). There is no contradiction
however, since these exponentially growing wavepacket 
solutions cannot be Fourier transformed
in time, so need not be expressible as superpositions of 
time-independent mode solutions.} 
Since the negative frequency part of this wavepacket determines the number
of created particles in the final outgoing wavepacket, that number will 
grow exponentially in the time between the
initial surface $\Sigma$ and the emergence of the outgoing
wavepacket $\psi$. Viewed forward in time,
the Hawking effect is a self-amplifying process since
the negative energy partners of the Hawking particles 
return to the event horizon (in the form of a pair of $k_+$ and
$k_-$ packets) and stimulate the emission
of more radiation and more partners. The wavepacket trajectories
associated with this forward in time picture are shown in figure 
\ref{partner}. 
 
For a fermionic field, the above discussion is modified only
by the fact that all wavepackets have positive norm,
so equation (\ref{amp}) implies that 
$||-s_1||^2$ is {\it smaller} in magnitude than $||-s_2||^2$.
This means that the number of created particles will
be exponentially {\it damped} in time. In effect, the allowed
states between the horizons 
for the negative energy partners of the Hawking particles
become filled, cutting off further pair creation.

One further important point can be extracted from this analysis. 
Since a single particle/partner  
wavepacket pair evolves to a sequence of outgoing wavepackets
as shown in figure \ref{partner}, the
states of all these outgoing wavepackets will be {\it correlated}.
This can also be seen from the backwards in time picture. 
It is clear from figure \ref{instabgraph} that successive 
outgoing wavepackets
will have past histories that partly overlap, in particular on the
initial ground state surface, so there will be
correlations between the quanta emitted 
from the horizon at different times. These correlations
are in sharp contrast to the usual Hawking effect
which produces
uncorrelated thermal radiation. The information loss that is 
normally associated with the correlations between Hawking quanta
and their partners is largely eliminated, since an unending sequence of
Hawking quanta is coherently correlated to the same partner degrees 
of freedom.

\section{Quantitative analysis}
\label{quantitative}

The qualitative analysis of the previous section will now be sharpened
by  explicitly constructing the wavepacket solutions discussed there.
This will allow us to quantify the amount of amplification, suppression,
and correlation of the black hole radiation.
In the first two subsections
we treat only the bosonic case, and in the last subsection
we discuss the fermionic case.

\subsection{Wavepacket solutions}
\label{wpsolutions}

The basic idea applied here is to patch together local
wavepacket solutions with the aid of ``evolution formulae". 
The derivation of these evolution formulae
is discussed in the Appendix of this paper.
They are derived using connection formulae
for time-independent mode functions which are obtained by matching WKB
solutions to near-horizon approximations. Forming wavepackets 
with the mode functions we then obtain the evolution formulae 
for the wavepackets. 

\begin{figure}[hbt]
\centerline{
\psfig{figure=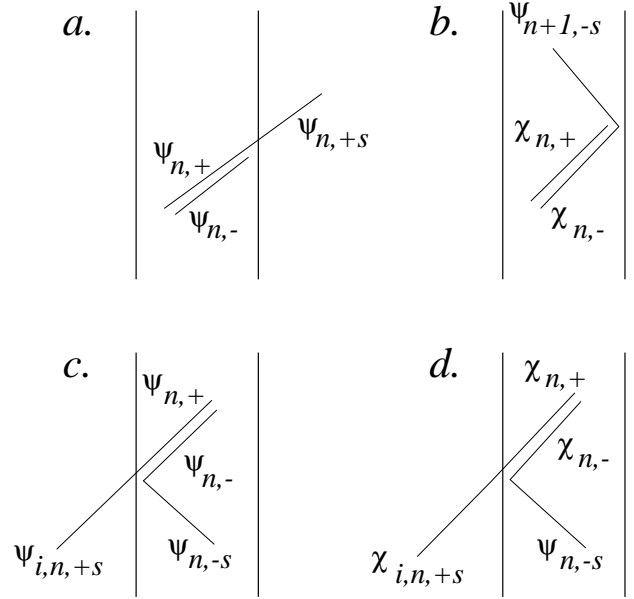,angle=-90,height=8cm}}
\caption{\small Spacetime sketches of the local wavepacket evolutions
given by (\ref{+s}-d) respectively.} 
\label{basicgraphs}
\end{figure}

Evolution formulae are needed for two different boundary conditions at both
the inner and outer horizons, corresponding to the spacetime diagrams
in figures 4a-d. Using the notation ``$f \rightarrow g$ " to denote
``$f$ evolves to $g$ " (forward in time), 
the evolution formulae about the outer horizon are: 
\begin{mathletters}
\begin{eqnarray}
\psi_{n,+} 
+ \psi_{n,-} 
\rightarrow &\psi_{n,+s} \label{+s}\\
\chi_{n,+} 
+ \chi_{n,-}
\rightarrow &\psi_{n+1,-s}, \label{-s} 
\end{eqnarray}
\text{while about the inner horizon they are}
\begin{eqnarray}
\psi_{i,n,+s}
+ \psi_{n,-s} 
\rightarrow &\psi_{n,+} + \psi_{n,-} \label{first} \\
\chi_{i,n,+s}
+ \psi_{n,-s} 
\rightarrow & \chi_{n,+} + \chi_{n,-}, \label{rest} 
\end{eqnarray}
\end{mathletters}
where all packets have been left unnormalized in order
to keep the formulae as simple as possible.
The evolution formulae given here are preferred for evolving packets
backwards in time.  Following the same techniques described
in the appendix evolution formulae more conducive to evolving
wavepackets forward in time can be derived.
The $+,-,+s$ and $-s$ subscripts
denote which type of wavevector the packet is peaked about.
$\psi_{n,+s}$ lies outside the outer horizon, while 
$\psi_{i,n,+s}$ and $\chi_{i,n,+s}$ both lie inside the inner horizon. 
The subscript $n$ is a sort of time variable. Translation of $n$ by one unit
has the effect of translating the wavepacket in time by a certain
amount and also distorting the wavepacket. 
Note that in the evolution formula 
at the outer horizon (\ref{-s}) 
$n$ increases by one unit on the $\psi_{n+1,-s}$ wavepacket. 
The evolution 
formulae (\ref{+s}-d) are basically scattering
solutions about a black hole event horizon with (\ref{+s}-b)
corresponding to scattering the $\psi_{n,+s}$ and $\psi_{n,-s}$ packets 
off the outer horizon backward in time (see figures \ref{basicgraphs}a
and b respectively) and (\ref{first}-d) corresponding 
to scattering the resulting
combinations of + and $-$ 
packets off the inner horizon 
backward in time as well (see figures \ref{basicgraphs}c and d respectively).

To construct the wavepacket solution with final data consisting
of a $k_{+s}$ packet outside the outer horizon 
we start with the local solution (\ref{+s}) (figure 
\ref{basicgraphs}a).
This clearly is not a global solution since $\psi_{n,\pm}$ do not
solve the equation of motion (\ref{eom}) about the inner horizon.
The combination may be replaced by (\ref{first}) which
does however.  This
results in the evolution formula 
\begin{equation}
(\psi_{i,n,+s} + \psi_{n,-s}) \rightarrow \psi_{n,+s}.
\label{laststep}
\end{equation}

The evolution (\ref{laststep}) is also 
not a global solution to the equation of motion (\ref{eom})
however since $\psi_{n,-s}$ is not a solution about the outer
horizon.  Using (\ref{-s}) to evolve $\psi_{n,-s}$ about the 
outer horizon, followed by (\ref{rest}) to evolve the
resulting wavepackets $\psi_{n-1,+}$ and  $\psi_{n-1,-}$ 
about the inner horizon, we obtain an evolution formula that 
can be iterated indefinitely, 
\begin{equation}
(\chi_{i,n-1,+s} + \psi_{n-1,-s}) \rightarrow \psi_{n,-s}.
\label{iterate}
\end{equation}
Beginning with ({\ref{laststep}) and iterating
(\ref{iterate}) $(n-m)$ times yields
\begin{eqnarray}
\biggl(\psi_{i,n,+s}  +  \sum^{n-m}_{j=1} \chi_{i,n-j,+s}
 + \psi_{m,-s}\biggr) \rightarrow \psi_{n,+s}.
\label{interevol}
\end{eqnarray}
In this manner we can evolve the final wavepacket packet back to the
spacelike surface where the initial ground state boundary condition is 
defined.
Solutions of the form (\ref{interevol})
correspond to those used in the qualitative
discussion of section \ref{wp}, which are depicted in 
Fig. \ref{instabgraph} (wherein the first $k_{-s}$ packet has been
traded for a $k_{\pm}$ pair using (\ref{-s}).)

If the horizons are not sufficiently widely separated then
intermediate wavepackets that arise between the
initial and final packets of (\ref{interevol}) will
overlap with the initial and final wavepackets, thus complicating
the analysis of particle creation. 
We can avoid such complications by constructing a different solution.
Setting $n$ equal to $m$ in (\ref{laststep})
and subtracting from (\ref{interevol}), we obtain
\begin{eqnarray}
\biggl(\psi_{i,n,+s} +
\sum^{n-m}_{j=1} \chi_{i,n-j,+s} 
&& - \psi_{i,m,+s} \biggr) \nonumber \\
&& \rightarrow \bigl(\psi_{n,+s} - \psi_{m,+s}\bigr). 
\label{twoout}
\end{eqnarray}
This solution corresponds to sending $(n-m)$ $\chi_{i,k,+s}$ packets and
a pair of $\psi_{i,k,+s}$ packets
into the inner horizon and getting a pair of $\psi_{+s}$ packets
out of the outer horizon.

\subsection{Particle creation}
\label{creationcomp}

We can now compute
the average number of particles in the wavepacket\footnote{We
use a $\hat{}$ to denote normalized wavepackets.}
$\hat{\psi}_{n,+s}$ and the correlations between emitted particles 
for different values of the time index $n$. 
To begin with let us evaluate the occupation number 
$\langle 0| N(\hat{\psi}_{0,+s})|0\rangle$ of the 
first outgoing packet after the initial state condition
is assumed.
Then the only evolution formula we need is (\ref{+s}),
with $n=0$.
The annihilation operator for a normalized wavepacket $f$ is given by 
$a(f)=(f,\widehat{\phi})$
(\ref{annih}). 
Taking the inner product of (\ref{+s}) with the quantum field
$\widehat \phi$, and using the ground
state conditions 
\beq
a(\hat{\psi}_{0,+})|0\rangle=0=
a(\hat{\psi}_{0,-}^*)|0\rangle, 
\label{ground1}
\eeq
we obtain
\begin{eqnarray}
\langle 0|  N(\hat{\psi}_{0,+s}) |0\rangle
&= &-  \frac{(\psi_{0,-}, \psi_{0,-})}{(\psi_{n,+s}, \psi_{n,+s})}
\nonumber\\
&= & \frac{1}{\omega_{u}  - \omega_{l}} \int_{\omega_l}^{\omega_u}
d\omega \,  \frac{1}{e^{2\pi\omega/\kappa} - 1}.
\label{N0explicit} 
\end{eqnarray}
(The norms (\ref{norms}) were used in the last equality.)
This is just the Planck distribution
at the Hawking temperature $T_H = \kappa/2 \pi$,
as was shown previously\cite{Unruhsuper,Corsuper} 
for a superuminal dispersive field theory 
in the case that there is just one horizon. 
It holds for $\kappa\lesssim\omega\ll k_0$.

We would have obtained a different
result for $\langle 0|  N(\hat{\psi}_{0,+s}) |0\rangle$
had we replaced the initial conditions (\ref{ground1}) with,
for example,
\beq
a(\hat{\psi}_{i,0,+s})|0\rangle = 0 =a(\hat{\psi}^{*}_{0,-s})|0\rangle.
\label{ground2}
\eeq
Indeed, from (\ref{laststep}) 
with $n=0$, we find that if (\ref{ground2}) holds
the occupation number of $\hat{\psi}_{0,+s}$
is given by 
\begin{eqnarray}
\langle 0|N(\hat{\psi}_{0,+s}) |0\rangle
 &=& -  \frac{(\psi_{0,-s}, \psi_{0,-s})}{(\psi_{0,+s}, \psi_{0,+s})}
\nonumber\\
&=& \frac{1}{\omega_{u}  - \omega_{l}} \int_{\omega_l}^{\omega_u}
d\omega \, \nonumber\\
&&~~\times \frac{2 \bigl(1 - \cos(\theta_+(\omega)-\theta_-(\omega))\bigr)}
{e^{2\pi\omega/\kappa} + e^{- 2\pi\omega/\kappa} - 2}. 
\label{altN0} 
\end{eqnarray}
The phase angles $\theta_{\pm}(\o)$ are defined implicitly
in (\ref{phase}), and
the norms (\ref{norms}) were used in the last equality
of (\ref{altN0}).
This differs from the thermal result (\ref{N0explicit}).

It is not yet clear to us what is the ``correct" initial condition on the
quantum state of the field. To determine this would require
following the evolution of the field state as 
the black hole (or condensed matter black hole analog)
forms. It does seem however that the conditions 
$a(\hat{\psi}_{i,k,+s})|0\ra=0$
are likely to hold, 
while the remaining specification of the state
remains to be determined. 
Fortunately these conditions alone suffice to determine the rate 
of growth of the number of particles emitted and the correlations between
them.

In order to find the number and correlations in the radiation
for $n>0$ we use the solution (\ref{twoout}).
We take $n>m\gg1$ so that the intermediate wavepackets that
entered the construction of (\ref{twoout}) will not have any
support on the initial surface, and we assume the 
ground state conditions 
\beq
a(\hat{\psi}_{i,k,+s})|0\rangle=0=a(\hat{\chi}_{i,k,+s})|0\rangle
\label{ground3}
\eeq
for $k\ge m$. 
Taking the inner product of (\ref{twoout})
with the field operator ${\widehat \phi}$ and using  
conditions (\ref{ground3}) we obtain
\begin{equation}
a(\hat{\psi}_{n,+s})|0\rangle = 
\sqrt{\frac{(\psi_{m,+s}, \psi_{m,+s}) }{ (\psi_{n,+s}, \psi_{n,+s})}}
a(\hat{\psi}_{m,+s})|0\rangle,
\end{equation}
from which it follows that 
\begin{eqnarray}
 &\langle& 0 | a^{\dag}(\hat{\psi}_{k,+s}) a(\hat{\psi}_{n,+s}) |0\rangle =
\nonumber \\  
 &~&   \frac{(\psi_{m,+s}, \psi_{m,+s}) }
{\sqrt{(\psi_{k,+s}, \psi_{k,+s}) (\psi_{n,+s}, \psi_{n,+s}) }} 
\, \langle 0| N(\hat{\psi}_{m,+s}) |0\rangle.
\label{corr2}
\end{eqnarray}
In particular taking $k=n$ we obtain
\begin{equation}
\langle 0| N(\hat{\psi}_{n,+s}) |0\rangle = 
 \frac{(\psi_{m,+s}, \psi_{m,+s})}
{(\psi_{n,+s}, \psi_{n,+s})} 
\, \langle 0| N(\hat{\psi}_{m,+s}) |0\rangle.
\label{Nn}
\end{equation}
The norm of $\psi_{n,+s}$
(\ref{norms}) is given by 
\begin{equation}
(\psi_{n,+s}, \psi_{n,+s})
= 4 \pi \int^{\omega_u}_{\omega_l} d\omega 
\left(1+ 
\frac{1-\cos(\theta_+ -\theta_-)}{2\sinh^2(\pi\omega/\kappa)}\right)^{-n}
\end{equation}
where we have used (\ref{Ts}). This
decreases monotonically with $n$
except for at most a discrete set of 
frequencies for which $\theta_+(\omega) = \theta_-(\omega) + 2 \pi k$
for some integer $k$. (For these frequencies $T_1$ (\ref{Ts}) vanishes,
so according to (\ref{min}) the corresponding mode is 
a bound state trapped between the horizons.) 
Therefore the particle creation in $\hat{\psi}_{n,+s}$
increases monotonically with $n$, diverging as $n\rightarrow\infty$.
In particular, 
if the wavepackets are narrowly peaked about a frequency $\omega$,
(\ref{Nn}) yields
\begin{equation}
\frac{\langle 0| N(\hat{\psi}_{n,+s}) |0\rangle}
{\langle 0| N(\hat{\psi}_{m,+s}) |0\rangle} = 
\left(1+ 
\frac{1-\cos(\theta_+ -\theta_-)}{2\sinh^2(\pi\omega/\kappa)}\right)^{n-m},
\end{equation}
which grows exponentially with $n-m$.

A measure of the correlation between emitted particles is given
by
\begin{eqnarray}
C(m,n) & := & \frac{\langle 0| a^{\dag}(\hat{\psi}_{m,+s}) 
a(\hat{\psi}_{n,+s}) |0\rangle}
{\Bigl(\langle 0| N(\hat{\psi}_{m,+s})|0\rangle
 \langle 0| N(\hat{\psi}_{n,+s})|0\rangle  \Bigr)^{1/2}}\nonumber \\ 
 &=& 1,
\label{C}
\end{eqnarray}
independent of the difference $n-m$.
This should be contrasted with the correlation obtained 
when ${\widehat \phi}$ satisfies the ordinary wave equation,
\beq
C(m,n)= (\hat{\psi}_{m,+s}, \hat{\psi}_{n,+s}),
\label{Cord}
\eeq
which is nonvanishing only to the extent that these 
wavepackets are not orthogonal. 
As $n-m$ grows the overlap of these wavepackets and hence 
the correlation (\ref{Cord}) goes to zero, whereas the 
correlation (\ref{C}) remains.

Finally let us estimate 
the time between the successive particle emissions (see
figure \ref{partner}), i.e, the difference in times when successive 
$\psi_{n,+s}$ packets (\ref{packets}) cross a fixed 
$x$ coordinate.  The trajectory of the
packets is given approximately by the condition of stationary phase, 
\begin{equation}
t =\frac{d}{d\omega}\Arg\Bigl(A_n(\o)\phi_{+s}(x,\o)\Bigr)
\end{equation}
and therefore the time $\Delta t$ 
between the $n^{th}$ and $(n+1)^{th}$
packets crossing the coordinate $x$ is given approximately by
\begin{eqnarray}
\Delta t &\approx&  \frac{d}{d\omega}\Arg\left(T_3\right)\nonumber\\
&=&\frac{d}{d\omega} \Biggl(\gamma +
\Arg\left(e^{-\pi\omega/\kappa}e^{-i \theta_+} - 
e^{\pi\omega/\kappa}e^{-i \theta_-}\right)\Biggr)
\end{eqnarray}
where we have substituted for $A_n$ using (\ref{packetcoeffs}),
and $T_3$ is given by (\ref{Ts}). 

Using the results given in the appendix the $\o$-dependence
of the phase factors $\gamma(\o)$
and $\theta_{\pm}(\o)$ can be computed. 
Rather than carrying out
this calculation---which we can only do explicitly for
any particular $v(x)$ in some
approximation anyway---let us make a rough estimate. The
interval $\Delta t$ is determined by the time it takes
a wavepacket to ``bounce" back and forth between the horizons.
If $v(x)+1$ is of order unity between the horizons, then
using the group velocity of the $k_{-s}$ and $k_{\pm}$ waves
one finds that this bounce time is of order $a$, the coordinate
distance between the horizons.

\subsection{Fermionic case} 
\label{fermions}

In this section we briefly describe the differences in the
quantitative analysis of the fermion case.
The derivation of the wavepacket solutions for fermions parallels
that given for the scalar field in section \ref{wpsolutions}
and yields an evolution formula very 
similar to (\ref{twoout}). The final expression for
the average value of the number operator is identical in form
to (\ref{Nn}), however the norm $N_{n,+s}$ now increases monotonically
with $n$ so that the number expectation value decreases
exponentially in $n$.  This is to be expected since 
unlike the scalar case, the conserved norm $\la f,f\ra$ (\ref{fip}) 
is positive definite.
The effect of this crucial difference is that instead of
exponential growth of Hawking radiation we now get
exponential decay.

\section{Discussion}
\label{discussion}
We left the question of the ``correct" initial condition unanswered.
For a condensed matter black hole it should be straightforward 
to deduce this by following the state of the field as
the ``black hole" forms.
It seems fairly clear that the $\psi_{i,+s}$
wavepackets inside the inner horizon will be in their ground states.
What is less clear is the state of the wavepackets $\psi_+$,
$\psi_-^*$ and $\psi_{-s}^*$ between the horizons.
For a real black hole---if one wants to entertain the possibility
of superluminal dispersion---the same may be true. The $\psi_{i,+s}$
wavepackets inside the inner horizon arise from ingoing waves that
scatter around or through the central 
singularity of a Reissner-Nordstr\"om black hole in the manner
discussed in \cite{Jacsemi}. Since the region inside the
inner horizon is static, it would seem plausible
that these are in their ground state as well.

Another issue we have not touched is that of the 
gravitational back-reaction
to the radiation studied here. In the bosonic case the exponential 
growth of the number of negative energy Hawking partners between the
horizons would surely rapidly entail a strong gravitational reaction.
In the fermionic case, the exponential suppression of radiation
leads quickly to a state with no radiation at all. This is hard
to reconcile with the usual picture in which Hawking radiation is 
a robust consequence of a general ``well-behaved" state near the
horizon. One would expect that although the negative energy states
of the Hawking partners in the ergoregion between the horizons 
become filled, there is not all that much energy in these states
(since the partners at late times are the same as the partners
at earlier times due to the ``bounce" between the horizons)
so the back-reaction should be limited. If so, then why doesn't
the Hawking radiation continue?  The answer, it would seem, is that
although the state is reasonably well-behaved in terms of energy
density, it has peculiar features in just those modes relevant
to the Hawking effect.

\section*{Acknowledgements}
This work was supported in part by NSF grants PHY94-13253
and PHY98-00967 at the University of Maryland, and  
by the Natural Sciences and Engineering Research Council of Canada
at the University of Alberta. We are grateful to Renaud Parentani
for pointing out the limited applicability of some approximations in
a previous draft of this paper.

\appendix 
\section*{Wavepacket solutions}
\label{appen}

In this appendix we explain how the evolution formulae 
(\ref{+s}-d)
for wavepacket solutions are derived with the help of the
results of \cite{Corsuper}.
We treat only the bosonic case, although the fermionic case
is essentially identical. 

The wavepacket evolution formulae are inferred from connection
formulae for mode solutions to the field equation (\ref{eom}) of the form
\begin{equation}
u(t,x) = e^{- i \omega t} \phi(x,\omega),
\end{equation}
where
$\phi(x,\omega)$ satisfies the ordinary differential equation (ODE)
\begin{eqnarray}
 - \phi^{(iv)}(x) &&+ (1 - v^2(x))\phi^{\prime \prime}(x) 
 +  2 v(x) (i \omega - v^{\prime}(x)) \phi^{\prime}(x) \nonumber \\
&&- i \omega ( i \omega - v^{\prime}(x)) \phi(x) = 0.
\label{ODE}
\end{eqnarray}
In \cite{Corsuper} such solutions were constructed for a black hole
spacetime with a single horizon.  
The basic technique used was to find approximate solutions
to (\ref{ODE}) using the WKB approximation away from the horizon, and
to match these solutions across the horizon by comparing
to the near horizon solution obtained by the method of Laplace transforms. 

\subsection{Outer horizon connection formulae}
Assuming that the
horizon is located at $x = 0$, and that the metric in the vicinity
of the horizon is given by
\begin{equation}
v(x) \approx -1 + \kappa x,
\label{localv}
\end{equation}
the analysis of \cite{Corsuper} leads to the following two connection 
formulae:\ 
\begin{mathletters}
\begin{eqnarray}
  K (e^{\pi \omega/2\kappa}\phi_+ +  e^{-\pi \omega/2\kappa}\phi_-) 
& \leftrightarrow & \phi_{+s} \label{+smode} \\
- \phi_{-s} +
 K (e^{-\pi \omega/2\kappa}\phi_+ 
+  e^{\pi \omega/2\kappa} \phi_-) 
& \leftrightarrow & 0, \label{-smode} 
\end{eqnarray}
\end{mathletters}
where 
\begin{equation}
K=(\omega/2\sinh(\pi\omega/\kappa))^{1/2}.
\end{equation}
The notation
``$\phi_1(x) \leftrightarrow \phi_2(x)$'' denotes that 
the approximate
WKB solution $\phi_1(x)$  {\rm behind} the horizon connects
to the approximate
WKB solution $\phi_2(x)$ {\rm outside} the horizon.
The modes $\phi_{\pm},\phi_{\pm s}$ are approximate WKB solutions to
(\ref{ODE}) and are given by\footnote{We
have changed notation slightly from that in \cite{Corsuper}.  
We have added the lower
limit of integration $\pm \epsilon$ to the integrals 
appearing in the exponents 
and consequently the coefficients $C_{\pm, \pm s}$ 
acquire some $\epsilon$ dependence to compensate. 
Furthermore a factor of $i$  
appearing in the matching formulae of \cite{Corsuper}
has been absorbed in $\phi_-$ and the 
phase of $N$ as defined in \cite{Corsuper} has been absorbed 
into $\phi_{\pm s}$.  We
have also renamed the $\phi_{-m}$ solution in \cite{Corsuper}
as $\phi_{-s}$ here.}
\begin{eqnarray}
\phi_{\pm}(x)  \approx  && C_{\pm}(v(x)^2-1)^{-3/4} \nonumber \\
&& \times \exp\left(i\int_{-\epsilon}^{x}ds\, k_{\pm}(v(s),\omega)\right)
\label{pmWKB} 
\end{eqnarray}
\begin{equation}
\phi_{\pm s}(x)  \approx   C_{\pm s} 
\exp\left(i\int_{-\epsilon}^{x}ds\, k_{\pm s}(v(s),\omega)\right),
\label{spmWKB} 
\end{equation}
where the approximate WKB wavevectors are given by 
\begin{eqnarray}
k_{\pm} & \approx & \pm k_0\sqrt{v^2-1} 
+  \omega v/(1-v^2)\label{kwkbpm}\\
k_{\pm s} & \approx & \omega /(1+v),\label{kwkbspm}
\end{eqnarray}
provided we assume $\omega\ll k_0$ and choose 
$|x|, \epsilon \gg (\omega/k_0)^{2/3}/\kappa$. 
The WKB solutions are
only valid for $|x| \gg \kappa^{-1/3}k_0^{-2/3}$.  
The coefficients $C_{\pm, \pm s}$ are necessary to match
these WKB solutions to the near-horizon Laplace transform solutions.
They can be determined by comparing the Laplace transform solutions
given in \cite{Corsuper} to the matching formulae (\ref{+smode}-b)
with the WKB modes (\ref{pmWKB},\ref{spmWKB}) evaluated in the
small $x$ limit.\footnote{$x$ cannot be arbitrarily small however 
because the WKB
approximation breaks down as $x \rightarrow 0$.  In \cite{Corsuper}
it was shown that the WKB and Laplace transform approximate solutions
are both valid  when $\kappa^{-1/3}k_0^{-2/3} \ll |x| \ll \kappa^{-1}$, 
and therefore the matching can be done in this
range.  We also choose $\epsilon$ to satisfy the same
inequality.} We find that the coefficients are given by
\begin{eqnarray}
C_{\pm} & = & i^{(1 \mp 1)/2} \exp\Bigl(\mp i\frac{2}{3} \sqrt{2 \kappa/k_0} 
(k_0\epsilon)^{3/2}\Bigr) \nonumber \\ 
& \times & \exp\left(-i \frac{\omega}{2 \kappa} \ln(2 \kappa \epsilon)\right)
 \label{Apm} 
\end{eqnarray}
\beq
C_{\pm s}  =  \exp\biggl(i\frac{\omega}{\kappa} \ln(k_0\epsilon) 
+ i \frac{\pi}{4}  
 -  i\, {\rm arg}\Bigl(\Gamma(1+i\omega/\kappa)\Bigr)\biggr).\label{Apms}
\eeq

\subsection{Inner horizon connection formulae}
In the case of a black hole with both inner and outer 
horizons, the connection formulae (\ref{+smode}-b)  
remain valid
locally about each horizon (after some slight modifications to be
discussed presently), but the solutions are no longer global.
Assume that
the outer horizon is located at $x_o = 0$, with $v(x)$ taking the
same form as given by (\ref{localv}), and that the inner horizon
is at $x_i = -a$ with $v(x)$ near the inner horizon taking the
form
\begin{equation}
v(x) \approx -1  - \kappa (x + a).
\label{localvinner}
\end{equation}
(We assume that the surface gravity of the inner horizon has the same
magnitude as that of the outer horizon to simplify the results.  There is
no difficulty however in allowing the surface gravities to be different.) 
Then the connection formulae (\ref{+smode}-b) are valid for
equation (\ref{ODE}) locally about the outer horizon, where
the notation ``$\phi_1(x) \leftrightarrow \phi_2(x)$'' now denotes that
$\phi_1(x)$ is valid between the horizons and $\phi_2(x)$ is valid outside
the outer horizon.  

To find the ``local'' mode solutions about the inner horizon we 
reexpress the mode equation (\ref{ODE}) in terms of the new 
coordinate $y := -(x + a)$. The resulting $y$-equation is the complex
conjugate of the $x$-equation (\ref{ODE}), with $v(x)$ replaced by 
$\widetilde{v}(y):=v(-y-a)$. We denote the WKB mode solutions to this
$y$-equation by $\widetilde{\phi}_{\pm}$ and $\widetilde{\phi}_{\pm s}$.
Since we have chosen the surface gravities to have the same magnitude,
$\widetilde{v}(y)$ takes the same form near the inner horizon 
as $v(x)$ does near the outer horizon (\ref{localv}). 
Therefore the mode solutions near the inner horizon are the
complex conjugates of those about the outer horizon, with 
$x$ replaced by $y$, so they satisfy the same connection 
formulae: 
\begin{mathletters}
\begin{eqnarray}
 K \bigl(e^{\pi \omega/2\kappa}\widetilde{\phi}_+ 
+  e^{-\pi \omega/2\kappa}\widetilde{\phi}_-\bigr) 
& \leftrightarrow & \widetilde{\phi}_{i,+s} \label{+mode} \\
- \widetilde{\phi}_{-s}+
  K \bigl(e^{-\pi \omega/2\kappa}\widetilde{\phi}_+ 
+  e^{\pi \omega/2\kappa} \widetilde{\phi}_-\bigr) 
 & \leftrightarrow & 0. \label{-mode} 
\end{eqnarray}
\end{mathletters}
The notation ``$\widetilde{\phi}_1(y)
\leftrightarrow \widetilde{\phi}_2(y)$'' now denotes that 
$\widetilde{\phi}_1(y)$ is valid between the
horizons and $\widetilde{\phi}_2(y)$ is valid inside the inner horizon.  
We have included a subscript ``$i$'' in $\widetilde{\phi}_{i,+s}(y)$ 
to make it clear that this solution is valid only inside the inner horizon.
 
The WKB solutions $\widetilde{\phi}_{\pm,-s}(y(x))$ are in fact 
the same functions of $x$, up to $\omega$-dependent
phases, as the WKB solutions $\phi_{\pm,-s}(x)$
respectively.  To see this, note that the WKB modes $\phi_{\pm,-s}(x)$
given in (\ref{pmWKB},\ref{spmWKB}) all take the
form\footnote{We have dropped the $\pm,-s$ subscripts on
$\phi(x), C, f$, and $k$.}
\begin{equation}
\phi(x) = C(\omega) f(v(x))
\exp\left(i\int_{-\epsilon}^{x}ds\, k(v(s),\omega)\right)
\label{WKBouter}
\end{equation} 
where 
$C$ is $x$-independent and $f$ and $k$ are real functions of
$v(x)$. Since, as discussed above, the $x$ and $y$ equations 
are related by substituting 
$v(x)\rightarrow \widetilde{v}(y)=v(-y-a)$ and complex conjugating,
the WKB modes $\widetilde{\phi}_{\pm,-s}(y)$ are given by 
\begin{equation}
\widetilde{\phi}(y) = C^*(\omega) f(\widetilde{v}(y))
\exp\left(-i\int_{-\epsilon}^{y}ds\, k(\widetilde{v}(s),\omega)\right)
\label{WKBinner1}
\end{equation} 
Using $y=-x-a$ and $\widetilde{v}(y)=v(x)$, and 
changing the integration variable in (\ref{WKBinner1})
to $s'=-s-a$, we obtain  
\begin{equation}
\widetilde{\phi}(-x-a) = C^*(\omega) f(v(x))
\exp\left(i\int_{-a+\epsilon}^{x}ds'\, k(v(s'),\omega)\right),
\label{WKBinner2}
\end{equation}
which differs from (\ref{WKBouter}) only by an $\omega$-dependent
phase factor, i.e.,
\begin{equation}
\widetilde{\phi}(-x-a) = \Biggl(\frac{C^*(\omega)}{C(\omega)} 
\exp\left(i\int_{-a+\epsilon}^{-\epsilon}ds\, k(v(s),\omega)\Biggr)
\right)\, \phi(x).
\label{innerouter}
\end{equation}
We shall not attempt to compute these phase factors, but rather
shall assume the generic form
\begin{eqnarray}
\widetilde{\phi}_{\pm}(-x-a) &=& e^{i \theta_{\pm}(\omega)} 
\phi_{\pm}(x), \nonumber \\  
\widetilde{\phi}_{-s}(-x-a) &=& e^{i \gamma(\omega)} \phi_{-s}(x),
\label{phase}
\end{eqnarray}
where $\theta_{\pm},\gamma$ also depend on $v(x), k_0$ and
$a$ (the coordinate distance between the horizons) but
do not depend on
$\epsilon$ as long as it is chosen large enough so that the
WKB approximation holds.

Using the phase relations (\ref{phase}) the inner horizon
connection formulae (\ref{+mode},\ref{-mode}) 
can be reexpressed in terms of the same 
linear combinations of $\phi_{\pm}$ appearing in the
outer horizon formulae (\ref{+smode},\ref{-smode}):
\beq
K(\ep \phi_+   +    \em \phi_-) +  
e^{i \gamma} T_1 \phi_{-s} \\  
\leftrightarrow  T_2 \widetilde{\phi}_{i,+s} 
\label{pin}
\eeq
\beq
K(\em \phi_+  +  \ep \phi_-) + T_3\phi_{-s}  
 \leftrightarrow 
T_1 \widetilde{\phi}_{i,+s}  
\label{min} 
\eeq
with
\begin{eqnarray}
T_1 & = & \o^{-1}K^2(e^{-i \theta_+} - e^{-i \theta_-})\nonumber \\
T_2 & = 
& \o^{-1}K^2(\ept e^{-i \theta_+} - \emt e^{-i \theta_-})\nonumber\\
T_3 & = & \o^{-1}K^2e^{i \gamma}
(\emt e^{-i \theta_+} - \ept e^{-i \theta_-}).\label{Ts}
\end{eqnarray}

\subsection{Wavepacket evolution formulae}
We can now form wavepackets from the modes and use the mode connection
formulae to obtain wavepacket evolution formuale.
To keep the latter simple, the $\omega$-dependence
of the coefficients in the connection formulae is built into
the definition of the wavepackets as follows.
Define
\begin{eqnarray}
\psi_{n,+s} & = & 
\int^{\o_u}_{\o_l} \frac{d\o}{\sqrt{\o}}
\, A_n\,   e^{-i \o t} \phi_{+s} \nonumber \\
\psi_{n,-s}  & = & 
\int^{\o_u}_{\o_l} \frac{d\o}{\sqrt{\o}}
\, B_n\,   e^{-i \o t} \phi_{-s} \nonumber\\
\psi_{n,\pm}  & = & 
\int^{\o_u}_{\o_l} \frac{d\o}{\sqrt{\o}}
\, C_{n,\pm}\,   e^{-i \o t} \phi_{\pm}\nonumber \\
\chi_{n-1,\pm} & = & 
\int^{\o_u}_{\o_l} \frac{d\o}{\sqrt{\o}}
\, D_{n-1,\pm}\,   e^{-i \o t} \phi_{\pm}  \nonumber\\
\psi_{i,n,+s}  & = & 
\int^{\o_u}_{\o_l} \frac{d\o}{\sqrt{\o}}
\, E_n \,  e^{-i \o t} \widetilde{\phi}_{i,+s}\nonumber \\
\chi_{i,n-1,+s}  & = & 
\int^{\o_u}_{\o_l} \frac{d\o}{\sqrt{\o}}
\, F_{n-1} \,  e^{-i \o t} \widetilde{\phi}_{i,+s},  
\label{packets}
\end{eqnarray}
where the coefficients $A\dots F$ depend on $\o$,
and the mode functions $\phi$ depend on both $\o$ and $x$.

With these definitions the evolution formulae about the outer
horizon (\ref{+s},\ref{-s}) follow immediately from the outer horizon
connection formulae (\ref{+smode},\ref{-smode}) 
provided we choose $C_{n\pm}=K\exp(\pm\pi\o/2\k)\, A_n$
and $D_{n,\pm}=-K\exp(\mp\pi\o/2\k)\, B_{n+1}$.
Similarly, the inner horizon evolution formula
(\ref{first}) follows provided $B_n=e^{i\gamma}T_1 A_n$
and $E_n=T_2A_n$, while (\ref{rest}) requires 
$F_n=-T_1 B_{n+1}$ and $B_n=-T_3B_{n+1}$. The solution 
(up to an undetermined overall constant) is given by 
\begin{eqnarray}
A_n &=& (-T_3)^{-n}\nonumber \\
B_n &=& e^{i\gamma} T_1 (-T_3)^{-n}\nonumber\\
C_{n,\pm} &=& Ke^{\pm\pi\o/2\k}(-T_3)^{-n}\nonumber \\
D_{n,\pm} &=& -Ke^{\mp\pi\o/2\k}e^{i\gamma}T_1 (-T_3)^{-n-1}\nonumber\\
E_n &=& T_2 (-T_3)^{-n}\nonumber\\
F_{n} &=& -e^{i\gamma} (T_1)^2 (-T_3)^{-n-1}  
\label{packetcoeffs}
\end{eqnarray}
 
\subsection{Norm of the wavepackets}
The wavepackets defined in (\ref{packets}) are not
normalized. Their norms can be determined as follows.
A generic one of these wavepackets has the form
\beq
\psi = \int^{\o_u}_{\o_l} \frac{d\o}{\sqrt{\o}}
\,  G\,  e^{-i \o t} \phi,
\eeq
which has the norm (cf. (\ref{inner}))
\begin{eqnarray}
(\psi,\psi)=
i\int dx\int\frac{d\o}{\sqrt{\o}}\int\frac{d\o'}{\sqrt{\o'}}
\Biggl\{G^*_\o G_{\o'}e^{i(\o-\o')t}  \nonumber\\
 \Bigl[\phi^*_\o(\partial_t+v\partial_x)\phi_\o'-
\phi_\o'(\partial_t+v\partial_x)\phi^*_\o \Bigr]\Biggr\}.
\label{normform}
\end{eqnarray}
The norm is conserved under time evolution so it suffices
to evaluate it at any one time. 

The key assumption we
need in order to evaluate the norm is that at some time the 
wavepacket is confined to a constant $v$ region. 
This is certainly the case for the $+s$ wavepackets,
since they are outgoing and eventually reach the asymptotic
region. If the region between the horizons is large and
has a large constant velocity region, then it may similarly
hold for the $-s$ and $\pm$ wavepackets as well. Alternatively,
these wavepackets  spend some time squeezed near the horizon,
with wavelengths much smaller than the length scale over which
$v(x)$ changes (but, in the case of $k_s$, still much longer than
$k_0^{-1}$, so we can nevertheless use the small $k$ approximation).
If the wavepacket is contained in a constant $v$ region
then, for
the purposes of evaluating the norm, we can imagine this 
region to extend to infinity in both directions. The
fixed $\o$ mode equation (\ref{ODE}) 
in a constant $v$ region has solutions
$\phi_\o=C_\o \exp(ikx)$, where $\o-vk=\pm F(k)$ with $F(k)$
given by (\ref{F}). Matching to the WKB modes 
(\ref{pmWKB},\ref{spmWKB})
we see that $|C_\o|=\{1,(v^2-1)^{-3/4}\}$ for the $\pm s$ modes and 
$\pm$ modes respectively.
Thus we have 
\beq
\int dx \phi^*_{\o'} \phi_{\o} = 
 2\pi \{1,(v^2-1)^{-3/4}\} \left|\frac{d\o}{dk}\right| 
\delta (\o'- \o). 
\label{modeoverlap}
\eeq
Using (\ref{modeoverlap}) in (\ref{normform}) yields
\beq
(\psi,\psi)=
4\pi \{1,(v^2-1)^{-3/2}\}  
\int d\o \, |G_\o|^2 \frac{\pm F}{\o}\left|\frac{d\o}{dk}\right|.
\label{norm1}
\eeq
Using the small and large $k$ approximations for 
$k_{\pm s}$ and $k_{\pm}$ respectively, we find that
\beq
 \{1,(v^2-1)^{-3/2}\}\frac{\pm F}{\o}\left|\frac{d\o}{dk}\right|
=\pm \{1 ,\o^{-1} \}
\label{id}
\eeq
respectively. Thus, finally, 
\beq
(\psi,\psi)=
\pm 4\pi 
\int d\o \, |G_\o|^2 \{1 ,\o^{-1}\} .
\label{norm2}
\eeq

With (\ref{norm2}) and the coefficients (\ref{packetcoeffs})
for the wavepackets (\ref{packets}) we obtain the norms
needed in section \ref{creationcomp}:
\begin{eqnarray}
(\psi_{0,-}, \psi_{0,-})
&=& - 4 \pi \int^{\omega_u}_{\omega_l} d\omega \,
\frac{1}{e^{2 \pi \omega/\kappa} -1}, 
\nonumber\\
(\psi_{n,+s}, \psi_{n,+s})
&=&  4 \pi \int^{\omega_u}_{\omega_l} d\omega \,
|T_3|^{-2n},\nonumber \\
(\psi_{0,-s}, \psi_{0,-s})
&=& - 4 \pi \int^{\omega_u}_{\omega_l} d\omega \,
|T_1|^2\nonumber\\
&=&- 4 \pi \int^{\omega_u}_{\omega_l} d\omega \,
\frac{2 \bigl(1 - \cos(\theta_+(\omega)-\theta_-(\omega))\bigr)}
{e^{2\pi\omega/\kappa} + e^{- 2\pi\omega/\kappa} - 2}.\nonumber\\
\label{norms}
\end{eqnarray}

\end{document}